\documentclass[a4paper,11pt]{article}
\usepackage{soul}
\usepackage[usenames,dvipsnames]{color}
\usepackage{jheppub}
\usepackage{esvect}
\usepackage{amsmath, amssymb, slashed, epsf, color, graphicx, latexsym, tensor, physics, xcolor}
\usepackage{epsfig}
\usepackage{comment}
\usepackage{graphics}
\usepackage{mathtools}
\usepackage{appendix}
\usepackage{inputenc}
\usepackage{lmodern}
\usepackage{indentfirst}
\usepackage{leftindex}
\newcommand{\e}[1]{\text{e}^{#1}}

\title{Logarithmic corrections for near-extremal black holes}
\author{Nabamita Banerjee,}
\author{Muktajyoti Saha,}
\author{and Suthanth Srinivasan}

\affiliation{Indian Institute of Science Education and Research Bhopal,\\ Bhopal Bypass, Bhauri, Bhopal 462066, India}

\emailAdd{nabamita@iiserb.ac.in}
\emailAdd{muktajyoti17@iiserb.ac.in}
\emailAdd{suthanth23@iiserb.ac.in}

\abstract{We present the computation of logarithmic corrections to near-extremal black hole entropy from one-loop Euclidean gravity path integral around the near-horizon geometry. We extract these corrections employing a suitably modified heat kernel method, where the near-extremal near-horizon geometry is treated as a perturbation around the extremal near-horizon geometry. Using this method we compute the logarithmic corrections to non-rotating solutions in four dimensional Einstein-Maxwell and $\mathcal{N} = 2,4,8$ supergravity theories. We also discuss the limit that suitably recovers the extremal black hole results. }

\begin{document}

\maketitle

\section{Introduction}
\noindent The universal structure of black hole entropy is a powerful property of quantum gravitational theories. In the semiclassical regime, the entropy has a universal form, proportional to the area of the horizon \cite{Bekenstein:1973ur, Hawking:1976de}. The leading-order quantum correction to this area law depends on the logarithm of horizon size \cite{Solodukhin:1994yz, Mann:1997hm, Medved:2004eh, Cai:2009ua, Aros:2010jb}. These logarithmic corrections are semi-universal in nature, in the sense that they depend only on the infrared data of the theory. Euclidean gravity formalism \cite{Gibbons:1976ue, York:1986it} has been proven successful in the computation of black hole entropy. The area law can be reproduced from a saddle point approximation of the Euclidean path integral whereas the one-loop contributions to the path integral capture the logarithmic corrections. A complete microscopic description of black hole entropy is however not understood yet, which requires the UV completion of gravity theories. Nevertheless, any sensible UV-complete theory of gravity should correctly reproduce the area and logarithmic terms in black hole entropy. \\

\noindent Generic black hole solutions can be categorized as extremal and non-extremal, depending on whether their temperature is zero or non-zero respectively. In the near-horizon region of an extremal black hole, an infinitely long AdS$_2$ factor emerges. This results in an enhancement of symmetries which in turn govern the dynamics of such black holes. This feature does not hold for non-extremal black holes, where the full geometry is required to understand their dynamics. In generic theories, the Bekenstein-Hawking entropy of extremal black holes can be easily computed using Sen's entropy function formalism \cite{Sen:2005wa, Sen:2008yk, Sen:2007qy}. Beyond the semiclassical regime, the idea has also been generalized to a quantum entropy function which can be used to compute the logarithmic corrections for extremal black holes \cite{Sen:2008vm, Banerjee:2009af, Banerjee:2010qc, Karan:2019gyn, Banerjee:2020wbr}. These formulations depend on the emergent near-horizon AdS$_2$ factor and its symmetries. Sen and collaborators also developed a technique to extract the logarithmic contributions to the entropy of generic (non)-extremal black holes using the heat kernel of the one-loop action of a gravity theory. For extremal black holes, again the near-horizon geometry was used \cite{Banerjee:2011jp, Sen:2012cj, Sen:2012dw, Bhattacharyya:2012wz}. Whereas for the non-extremal black holes, the results depend on the full geometry \cite{Sen:2012kpz}\footnote{The technique was applied to various theories \cite{Karan:2020njm, Karan:2021teq, Banerjee:2021pdy, Karan:2022dfy}.}. String theory has provided a microscopic realization of entropy for a class of supersymmetric extremal black holes \cite{Strominger:1996sh, Breckenridge:1996is, David:2006ru, David:2006ud, Sen:2008yk, Gupta:2008ki, Banerjee:2007ub, Banerjee:2008ky, Banerjee:2009uk, Banerjee:2009af}. For extremal black holes appearing in various string theories \cite{Banerjee:2010qc, Banerjee:2011jp, Sen:2012cj}, the matching of logarithmic corrections from gravitational and microscopic perspectives has been achieved. In the recent works \cite{H:2023qko, Anupam:2023yns}, the extremal black hole results are also correctly reproduced from the limit of a finite temperature geometry computation. \\

\noindent Black holes with very small temperatures are called \textit{near}-extremal. Such black holes with very small temperatures still exhibit some interesting features that are characteristic of extremal black holes. Thus they are good candidates to generalize the progress made for extremal black holes in the presence of small but non-zero temperatures. Recent studies have rendered near-extremal black holes to have quite distinct properties by their own virtue. Notably, it has been shown in \cite{Nayak:2018qej, Moitra:2019bub, Iliesiu:2020qvm} that the dynamics of such black holes can be obtained from an effective 1D Schwarzian description (See \cite{Heydeman:2020hhw, Kolekar:2018sba, Banerjee:2021vjy, Bhattacharyya:2023gvg} for generalizations to various systems). \cite{Iliesiu:2020qvm} shows that the low temperature dynamics is governed by one-loop corrections proportional to the logarithm of temperature, which are different than the usual logarithmic corrections. Recent works \cite{Iliesiu:2022onk, Banerjee:2023quv} trace back the origin of these terms from a 4D Euclidean gravity computation, in a language similar to that of the usual logarithmic contributions. It was shown that these terms appear from the one-loop quantization on the near-horizon near-extremal background, which is a small deviation of the extremal near-horizon geometry. The quantization procedure depends on the computation of the eigenvalues of the kinetic operator of small fluctuations around the classical background. The idea of \cite{Banerjee:2023quv} is to use first-order perturbation theory to compute these eigenvalues and extract the logarithm of temperature terms. \\

\noindent The goal of this work would be to analyze the logarithmic corrections for near-extremal black holes, carefully taking care of the differences with (non)-extremal 
solutions. Our approach would be to employ first-order perturbation theory \cite{Banerjee:2023quv} to correctly modify the usual heat-kernel method for this purpose. The usual approach for (non)-extremal black holes depends on a scaling property of parameters, where all the length scales are of the same order and thus scale uniformly. However, for near-extremal black holes, the issue is subtle as the inverse temperature and charges bring in large independent length scales. We first separate the logarithmic contributions coming from these two large parameters. We then use this method to compute the logarithmic corrections to near-extremal entropy in $\mathcal{N} = 2,4,8$ supergravity theories. We also recover the extremal logarithmic corrections for both supersymmetric and non-supersymmetric theories by appropriately taking the extremal limit of our results.
 Our modified heat-kernel method correctly reproduces the results for Einstein-Maxwell theory as earlier found in \cite{Iliesiu:2020qvm, Iliesiu:2022onk, Banerjee:2023quv}. We also find agreement with the existing results of \cite{Iliesiu:2022onk} for $\mathcal{N} = 2,4,8$ supergravity theories for a \textit{particular} near-extremal solution. However, to regulate a certain kind of divergence at a zero temperature limit of the near-extremal result, a sum over different saddle points was considered in \cite{Iliesiu:2022onk}. In our computation of the logarithmic correction for a non-rotating near-extremal black hole, such a regularization is not required. We elaborate on this issue further in the discussion section. \\

\noindent The paper is organized as follows: In section \ref{near-ext-sol}, we discuss the structure of the near-horizon geometry of a near-extremal black hole. In section \ref{heat-kernel}, we first review the heat kernel method to compute logarithmic corrections for extremal black holes and then discuss how to modify the method for computations for near-extremal black holes. This prescription is one of the main results of this paper. We also discuss how to take an appropriate extremal limit of the near-extremal computation. In section \ref{N=2}, we compute the logarithmic corrections to near-extremal black hole entropy appearing in $\mathcal{N} = 2$ supergravity theory. Section \ref{N=4,8} contains the results for the same for $\mathcal{N} = 4,8$ supergravity theories. Finally, we summarize and discuss our results in section \ref{concl}.

\section{Near-extremal background} \label{near-ext-sol}
\noindent In this section, we discuss the properties of a near-extremal black hole solution near the horizon. We will begin by reviewing the classical geometry \cite{Iliesiu:2020qvm, Banerjee:2023quv} in Einstein-Maxwell theory and then extend the idea to generic theories. Let us consider the 4D Einstein-Maxwell theory with the Euclidean action,
\begin{align}
    \mathcal{S} = -\int d^4x \sqrt{g} (R - F^2) \label{EM-act}.
\end{align}
We have set the Newton constant to $1/16\pi$. Therefore, the action has the dimensions of length-squared. A spherically symmetric black hole solution in this theory is described by the Reissner-N\"{o}rdstrom geometry, which is parametrized by mass and charge,
\begin{align}
   & ds^2 = g_{AB}dx^A dx^B = f(r)dt^2 + \frac{dr^2}{f(r)} + r^2d\Omega^2, \quad f(r) = 1 - \frac{2M}{r} + \frac{Q^2}{r^2}, \\
   & A_t = i Q \left( \frac{1}{r_{+}} - \frac{1}{r} \right) , \quad F_{rt} = \frac{i Q}{r^2}.  \label{RN}
\end{align}
For $M > Q$, we have a non-extremal black hole that has a finite temperature. At extremality i.e. for $M = Q$, the solution has a horizon radius given by $Q$ and it has zero temperature. We are interested in a near-extremal black hole, which has the same charge but mass slightly greater than the extremal mass. We parametrize the solution with charge $Q$ and temperature $T$. Since it is a near-extremal black hole, we work in the regime $QT\ll 1$, which signifies that we are very close to extremality. The reason for fixing the charge is that finally, we want to compute the entropy in a microcanonical ensemble. The quantum entropy function formalism \cite{Sen:2005wa, Sen:2008yk, Sen:2007qy} for extremal black holes already produces the microcanonical entropy from the gravitational side. Therefore, keeping the charges fixed while introducing a temperature preserves this feature for the charges. In the final expression, an inverse Laplace transform of variables from inverse temperature to energy would directly give us the microcanonical entropy of the near-extremal black hole.\\

\noindent As we move close to the horizon of an extremal black hole, an AdS$_2$ factor emerges and this results in a symmetry enhancement. The near-horizon throat being very large, these symmetries govern the dynamics of (near)-extremal black holes. The near-horizon geometry \cite{Iliesiu:2022onk, Banerjee:2023quv} is given by $ { 
g_{AB} = g^0_{AB} + T g^{(c)}_{AB}, A_B = {A}^0_B + T A^{(c)}_B }$, such that $\{g^0,A^0\}$ denotes the extremal AdS$_2\times$S$^2$ geometry:
\begin{align}
     & {g}^0_{AB}dx^A dx^B = Q^2 (d\eta^2 + \sinh^2{\eta}d\theta^2) + Q^2 (d\psi^2 + \sin^2{\psi}d\varphi^2), \nonumber \\
     & {F}^0_{\mu\nu} = \frac{i}{Q}\varepsilon_{\mu\nu}, \quad {A}^0_{\theta} = i Q (\cosh{\eta} - 1). \label{ext-NH} 
\end{align}
The small temperature causes deviations from this geometry, given by
\begin{align}
     & g^{(c)}_{AB}dx^A dx^B = 4\pi Q^3 (2+\cosh{\eta})\tanh^2\left(\frac{\eta}{2}\right) (d\eta^2 - \sinh^2{\eta}d\theta^2) + 4\pi Q^3 \cosh{\eta} d\Omega^2, \nonumber\\
     & F^{(c)}_{\mu\nu} = - 4\pi i\cosh{\eta}\varepsilon_{\mu\nu}, \quad  A^{(c)}_{\theta} = -2\pi iQ^2\sinh^2{\eta}. \label{NE-corr}
\end{align}
The near-horizon coordinates range from $0<\eta<\eta_0$ and $0<\theta<2\pi$ such that the horizon is located at $\eta = 0$. We will denote the coordinates on AdS$_2$ by $x^\mu$ and the coordinates on S$^2$ by $x^i$. $\varepsilon_{\mu\nu}$ is the Levi-Civita tensor on AdS$_2$, with the non-zero component being ${\varepsilon_{\eta\theta} = Q^2 \sinh{\eta}}$. The near-horizon geometry is glued to the asymptotic geometry at the boundary located at $\eta = \eta_0$, which is very far away from the horizon. Due to a small temperature, the throat is large enough to capture the properties of the black hole. The quantum corrections are large in the near-horizon region but very much suppressed in the asymptotic region. It essentially motivates us to quantize the system on the near-horizon background. This is extensively studied in \cite{Iliesiu:2020qvm, Iliesiu:2022onk, Banerjee:2023quv}. \\

\noindent In generic theories, the extremal black hole solution can be parametrized by the charges corresponding to various gauge fields. Similar to the Reissner-N\"{o}rdstrom scenario, we will keep those charges fixed and introduce a temperature parameter to describe a near-extremal solution. This will again cause first-order temperature deviation from the extremal near-horizon geometry in a way that the flux corresponding to different gauge field strengths remains the same as the extremal black hole. Hence, schematically we write the near-horizon geometry as,
\begin{align}
    \bar{\Psi}(T,Q_i) = \bar{\Psi}^0 (Q_i) + T\bar{\Psi}^{(c)} (Q_i). \label{gen-NE-NH}
\end{align}
Here, $\Psi$ denotes all the fields in the theory and $Q_i$ denotes the charges that parametrize the extremal solution. The bar is to signify that we are considering a classical background. $\bar{\Psi}^0$ denotes the field content of the extremal near-horizon geometry, whereas $\bar{\Psi}^{(c)}$ denotes the corrections in presence of small temperature. The charge dependence in $\bar{\Psi}^{(c)}$ can be derived from that of $\bar{\Psi}^0$. The charges are of the same order i.e. $Q_i \sim a$, where $a$ characterizes the horizon size of the extremal black hole. The working regime is set by $a T\ll 1$.

\section{Heat kernel prescription for logarithmic corrections}\label{heat-kernel}
\noindent Let us consider a theory described by Euclidean action $\mathcal{S}$. We would like to consider the corresponding Euclidean path integral to one-loop order. For this, we turn on small fluctuations ($\Psi$) around a classical background ($\Bar{\Psi}$) and expand the action to quadratic order in fluctuations. The zeroth-order term gives the saddle point contribution and the first-order term vanishes since the background satisfies the equations of motion. Due to the quadratic term, the one-loop path integral can be expressed as a Gaussian integral of bosonic and fermionic fluctuations,
\begin{align}
    Z = \int \mathcal{D}\Psi\ \e{-\int d^4x\sqrt{\bar{g}}\Psi^i\bar{\Delta}_{ij}\Psi^j}.
\end{align}
Here, the operator $\bar{\Delta}$ depends on the theory in consideration and can be obtained on any arbitrary background. The $i,j$ indices capture any sort of spacetime or internal indices for the fields. For bosonic variables, the operator is a two-derivative operator whereas for fermionic variables, it is linear in derivative. The one-loop path integral can in principle be obtained from the determinant of this operator $\bar{\Delta}$ on a particular classical background. The determinant may be ill-defined due to the presence of zero modes of the kinetic operator, but these are dealt with separately. The non-zero mode contributions can be expressed as,
\begin{align}
    \log{Z}_{\text{nz}} = -\frac{1}{2}\sideset{}{'}\sum_n\log{\kappa}_n + \frac{1}{4}\sideset{}{'}\sum_n\log{\kappa}^f_n.
    \label{logeigval}
\end{align}
The prime indicates that we are summing over non-zero modes only. We have written the bosonic and fermionic (denoted by a superscript $f$) contributions separately since the bosonic Gaussian integral gives $(\det\Bar{\Delta})^{-1/2}$ whereas the fermionic integral\footnote{We are considering Majorana fermions, which have half the degrees of freedom as compared to Dirac fermions due to reality constraint. Hence the Gaussian integral gives the square root of determinant.} gives $(\det\Bar{\Delta})^{1/2}$. Also, $\{\kappa_n^f\}$ denote the eigenvalues of the squared fermionic operator so that the eigenvalues $\kappa_n$ and $\kappa_n^f$ have the same dimensions. This is reflected by the additional factor of $1/2$ in front of the fermionic contribution.\\

\noindent Without directly computing the individual eigenvalues, the heat kernel corresponding to the kinetic operator can be used to evaluate \eqref{logeigval}. We work with the bosonic and fermionic parts separately. First, we will discuss the bosonic contributions and later draw an analogy for the fermionic case. The heat kernel for the kinetic operator of bosonic fluctuations is defined as,
\begin{align}
     K^{ij}(x, x'; s) = \sum_{n} \e{-\kappa_n s}f^i_n(x)f^j_n (x'). \label{heat-boson}
\end{align}
Here, $\kappa_n$ denotes the eigenvalues of the kinetic operator $\bar\Delta$, corresponding to the eigenfunction $f_n^i(x)$, which we are taking to be real. For complex eigenfunctions, we should take the complex conjugate for one of the eigenfunctions in the sum. The eigenfunctions are orthonormal with respect to the inner product,
\begin{align}
    \int d^4x\sqrt{\bar g} G_{ij}f^i_n(x)f^j_m (x) = \delta_{nm}. \label{ortho}
\end{align}
Here, $G_{ij}$ is the metric that is induced on the field space. The heat kernel satisfies a `heat equation' of the form $(\partial_s + \Bar{\Delta})K(x,x';s) = 0$, where we have suppressed the internal indices. For fermionic fluctuations, we will consider the heat kernel corresponding to the squared kinetic operator and further define the heat kernel with an additional factor of $-1/2$ such that\footnote{An important point to note is that we should only consider the eigenfunctions of the linear derivative kinetic operator and take the corresponding eigenvalues of the squared kinetic operators.},
\begin{align}
     K^{ij}_f(x, x'; s) = -\frac{1}{2}\sum_{n} \e{-\kappa_n^f s}f^i_n(x)f^j_n (x'). \label{heat-ferm}
\end{align}

\noindent To evaluate the logarithm of partition function \eqref{logeigval} in terms of the heat kernel, the following identity is used,
\begin{align}
    \log{(A\epsilon)} = -\int_{\epsilon}^{\infty} \frac{ds}{s}\e{-As}. \label{schwinger}
\end{align}
$\epsilon\to 0$ is a small cutoff introduced to regulate the integral, which in our case will be provided by the UV cutoff of the theory\footnote{In this integral, $A$ is considered to be much above the cutoff scale $\epsilon$. We will get back to this point in a later discussion.}. Using the above identity, the bosonic non-zero modes sector of \eqref{logeigval} can be expressed in terms of the trace of the heat kernel as,
\begin{align}
     \log{Z}_{\text{nz}} = \frac{1}{2}\sideset{}{'}\sum_n \int_{\epsilon}^\infty \frac{ds}{s} \text{e}^{-\kappa_n s} = \frac{1}{2}\int_{\epsilon}^{\infty} \frac{ds}{s}\int d^4 x \sqrt{g} K'(x, x; s). \label{logZ-trace}
\end{align}
Here, we are using the notation $K(x,x';s)\equiv G_{ij}K^{ij}(x,x'; s)$. Again, the prime denotes the non-zero contribution only. Since for fermionic fluctuations, the additional factor of $-1/2$ is absorbed into the definition of the heat kernel, the final form in terms of the trace remains the same. In particular, the fermionic non-zero mode contribution is given by,
\begin{align}
     \log{Z}_{\text{nz}} = -\frac{1}{4}\sideset{}{'}\sum_n \int_0^\infty \frac{ds}{s} \text{e}^{-\kappa^f_n s} = \frac{1}{2}\int_{\epsilon}^{\infty} \frac{ds}{s}\int d^4 x \sqrt{g} K'_f(x, x; s). 
\end{align}

\noindent In general, it is very difficult to compute the full one-loop contribution in an arbitrary theory. In \cite{Banerjee:2010qc, Banerjee:2011jp, Sen:2012cj, Bhattacharyya:2012wz, Sen:2012kpz, Sen:2012dw}, it was shown how to extract the logarithmic corrections to (non)-extremal black hole entropy from the heat kernel prescription. It was argued that the logarithmic corrections appear from the small $s$-region of the integration. Although the techniques for extremal and non-extremal black holes are essentially the same in spirit, there are some important differences between these two cases. One of them is that for the extremal black hole, the computation is done on the near-horizon background whereas for the non-extremal black hole, the full geometry must be used. The analysis for a near-extremal black hole will be a close cousin to the analysis for an extremal black hole, due to the presence of a large near-horizon throat. Also, the temperature and charge parameters introduce different scales in the solution in contrast with a generic non-extremal black hole where all the parameters have uniform scaling. Hence, we now briefly review the method for extremal black holes following the works of \cite{Banerjee:2011jp, Sen:2012kpz}. \\

\noindent We first consider the case of $\mathcal{N} = 2$ supergravity theory, which contains the Einstein-Maxwell theory as the bosonic sector. Here the horizon size is equal to the charge and the extremal black hole can be parametrized by charge only. The analysis of the subsequent part of this section can be easily generalized to arbitrary theories with various charges. Since the charges scale uniformly with the extremal horizon size, we can just work with one of the charges or the horizon size.

\subsection{Brief review of the  approach for extremal black holes}
\noindent In this section we briefly review the prescription for extremal black holes. Let us consider a spherically symmetric extremal black hole solution parametrized by its electric charge $Q$ only. Its near-horizon geometry\footnote{We will denote the extremal geometry with a superscript 0.} is AdS$_2\times$S$^2$ with radii $Q$. The extremal solution has the following scaling dependence,
\begin{align}
    g_{\mu\nu}^0 \sim Q^2, \quad A_{\mu}^0 \sim Q.
\end{align}
As mentioned earlier, the action must have the dimensions of length squared such that the quadratic fluctuated action should also satisfy this scaling property,
\begin{align}
    \mathcal{S}^{(1)}\equiv \int d^4x\sqrt{g^0}\Psi^i\Delta^0_{ij}\Psi^j \sim Q^2.
\end{align}
Due to the form of the extremal metric, in 4D we have $\sqrt{g^0}\sim Q^4$. Thus we find that the two-derivative kinetic operator must have the dependence $\sim Q^{-2}$ since it is constructed out of the extremal background. First, we consider the non-zero mode contribution to the partition function. From the scaling properties discussed above, we can find that the non-zero eigenvalues of the kinetic operator must be of the form,
\begin{align}
    \kappa^0_n \sim \frac{1}{Q^2}.
\end{align}
The dimensions of the eigenfunctions for different fields can be read off from the orthonormality condition \eqref{ortho}. The Schwinger parameter $s$ has dimensions of length squared. Using these dimensional dependencies, we perform a change of variable $s\to\hat{s} = s/Q^2$ such that \eqref{logZ-trace} becomes,
\begin{align}
    \log Z_{\text{nz}} = \frac{1}{2}\int_{\epsilon/Q^2}^{\infty} \frac{d\hat{s}}{\hat{s}}\int d^4 x \sqrt{g^0} K'(0; \hat{s}).
\end{align}
The hat on various quantities denote their dimensionless variants, after the $Q$-dependencies are stripped off. Due to the homogeneity of the extremal geometry, the coordinate dependence goes away from the trace of the heat kernel. As argued in \cite{Bhattacharyya:2012wz}, the logarithmic corrections appear from the small $s$-regime, where the heat kernel trace over a background can be expanded in terms of Seeley-de Witt coefficients, 
\begin{align}
    K(x,x;\hat{s}) = \sum_{m = 0}^\infty K_{-\frac{D}{2} + m}(x) \hat{s}^{-\frac{D}{2} + m}.
\end{align}
$K_{-\frac{D}{2} + m}(x)$ is a $2m$-derivative local diffeomorphism and gauge invariant term constructed out of the background fields. For extremal background, these coefficients are just constant terms \cite{Sen:2012kpz}. The $\hat{s}$-independent part of the trace, in particular, the integration of the coefficient $K_0$ determines the coefficient of the $\log{Q}$ term in $\log{Z_{\text{nz}}}$, given by,
\begin{align}
    \log{Z_{\text{nz}}}= \frac{1}{2}\int d^4 x \sqrt{g^0}\left( K_0 - K(0)\right)\log Q^2 = 8\pi^2 Q^4 (\cosh\eta_0 - 1)\left( K_0 - K(0)\right)\log Q.
\end{align}
Here, $K(0)$ determines the number of zero modes, given by,
\begin{align}
    N_{\text{zm}} = \int d^4 x \sqrt{g^0} K(0) = 8\pi^2 Q^4 (\cosh\eta_0 - 1) K(0); \quad K(0) \equiv \sum_{\text{zm}} G_{ij}f^i_n f^j_n. \label{Nzm}
\end{align}

\noindent Now we consider the contributions coming from the zero modes of the extremal black hole that arise in the near-horizon geometry. Zero modes are associated with the symmetries that are spontaneously broken by the black hole solution. Due to the presence of the AdS$_2$ boundary, the global symmetries of the solution get enhanced to the infinite-dimensional algebra of asymptotic symmetries. We have an infinite number of zero modes associated with the spontaneous breaking of these asymptotic symmetries. These fluctuations are generated by large gauge transformations and diffeomorphisms, which are non-normalizable\footnote{The asymptotic symmetries near null infinity for the black holes under consideration are the well-known BMS symmetries. For Einstein-Maxwell, Yang-Mills, and certain supersymmetric theories, the asymptotic symmetries are discussed in \cite{Fotopoulos:2019vac, Fan:2019emx, Fotopoulos:2020bqj, Banerjee:2021uxe, Banerjee:2022hpo, Banerjee:2022lnz}.}. There is no Gaussian suppression for these modes in the path integral, hence the integral is typically divergent. However, there can be non-trivial logarithmic corrections coming from the path integral measure of these modes and these contributions are fixed by the normalization conditions imposed on these modes.\\

\noindent To understand the contributions, we make a change of variables from the field variables to the large gauge transformation parameters that generate the corresponding fluctuations. These large gauge parameters do not depend on the charge, hence their integration does not give rise to any charge dependence. The non-triviality appears when the change of variables from the fluctuations to the gauge parameters is performed. The corresponding Jacobian due to the change of variable results in a particular dependence in charge. A factor of $Q^{\beta_r}$ is introduced in the measure from the Jacobian for each zero mode. The number $\beta_r$ depends on the particular type of mode in consideration. It can be fixed by analyzing the field fluctuations and the corresponding gauge parameters. We discuss this feature in the later sections when we consider extremal black hole solutions in particular theories. Hence, the zero mode contribution to the logarithm of partition function is given as,
\begin{align}
    \log Z_{\text{zm}} = \sum_r \beta_r N^r_{\text{zm}} \log{Q} = 8\pi^2 Q^4 (\cosh\eta_0 - 1)\sum_r \beta_r K^r(0) \log{Q}.
\end{align}
Here, we have included a factor of $Q^{\beta_r}$ for each type of zero mode, labeled by $r$, and the sum characterizes a sum over different types of zero modes. $N^r_{\text{zm}}$ denotes the number of such zero modes, given by \eqref{Nzm} where the sum should involve the zero mode eigenfunctions of that particular category. $\eta = \eta_0$ denotes the boundary of the near-horizon AdS$_2$ throat, which is very large. The boundary cutoff-dependent pieces in the expression of $\log{Z}$ can be interpreted as an infinite shift in the ground state energy. The cutoff-independent part will give the actual contribution to entropy.

\subsubsection*{Fermionic sector}
\noindent A similar analysis can be performed for the fermionic fluctuations also. Since we are considering the eigenvalues of the squared kinetic operator, the dimensional dependencies remain the same as that of the bosonic operator as both of these are two-derivative terms. The form of the non-zero mode contribution remains exactly the same i.e.
\begin{align}
    \log{Z_{\text{nz}}}= \frac{1}{2}\int d^4 x \sqrt{g^0}\left( K^f_0 - K^f(0)\right)\log Q^2.
\end{align}
Now we consider the zero modes contribution. The number of zero modes in terms of the trace of the heat kernel is given as,
\begin{align}
    N^f_{\text{zm}} = -2\int d^4 x \sqrt{g^0} K^f(0) = -16\pi^2 Q^4 (\cosh\eta_0 - 1) K^f(0); \quad K^f(0) \equiv -\frac{1}{2} \sum_{\text{zm}} G_{ij}f^i_n f^j_n. \label{ferm-no.}
\end{align}

\noindent The fermionic zero modes are spin-$3/2$ fluctuations, associated with the breaking of the supergroup of asymptotic symmetries and these are generated by non-normalizable spin-$1/2$ parameters. Since there is no Gaussian suppression, these integrals typically go to zero. To extract the charge dependence, we again perform a change of variable from the fields to the gauge parameters, which introduces a factor of $Q^{-\beta_f/2}$ per Majorana zero mode in the measure. Hence, the fermionic zero modes contribution to the logarithm of partition function is given as,
\begin{align}
    \log Z_{\text{zm}} = -\frac{1}{2}\sum_f \beta_f N^f_{\text{zm}} \log{Q} = 8\pi^2 Q^4 (\cosh\eta_0 - 1)\sum_r \beta_f K^f(0) \log{Q}.
\end{align}

\noindent The total logarithm of charge contribution to the extremal black hole partition function is given as,
\begin{align}
    & \log Z \sim (c_0 + c_0^f)\log Q, \\ 
    & c_0 = -8\pi^2 Q^4[K_0 + \sum_{r\in\text{zm}} (\beta_r - 1)K^r(0)], \\
    & c^f_0 = -8\pi^2 Q^4[K^f_0 + \sum_{f\in\text{zm}} (\beta_f - 1)K^f(0)]. \label{logQ-coeff-ext}
\end{align}
We will compare this coefficient with the corresponding coefficient for the near-extremal black hole.

\subsection{Approach for near-extremal black holes} \label{near-ext-approach}

\noindent In this section, we will consider the one-loop path integral around a spherically symmetric near-extremal black hole solution parametrized by temperature $T$ and charge $Q$ such that $QT\ll 1$. As discussed earlier, this solution also has a finite but very large near-horizon throat having a geometry that can be described as a linear-order temperature correction to the extremal AdS$_2\times$S$^2$ geometry. We will quantize the system on the near-horizon background following the approach of first-order perturbation theory \cite{Banerjee:2023quv}. However, we will resort to a heat kernel method that does not depend on explicit eigenvalue computations, in contrast with \cite{Banerjee:2023quv}. \\

\noindent The dogma of the first-order perturbation theory technique is that it recasts the quantization procedure around the near-extremal background into finding the spectrum (or in this case, the heat kernel) of a modified operator on the AdS$_2\times$S$^2$ geometry itself. Since the near-extremal background can be described as a linear order temperature deviation from the extremal one, the quadratic action can be rewritten as,
\begin{align}
    \mathcal{S}^{(1)} = -\int d^4x \sqrt{g^0}\ \Psi^i (\Delta^0_{ij} + T \Delta^{(c)}_{ij}) \Psi^j.
\end{align}
We have clubbed all the temperature-dependent corrections into the operator $\Delta^{(c)}$ and our interest lies in finding the heat kernel for the modified operator $\Delta \equiv \Delta^0 + T\Delta^{(c)}$. In particular, the eigenvalue equation of this operator is given by,
\begin{align}
    & \Delta_{ij} f^j_n = \kappa_n f_{n,i} , \quad \kappa_n = \kappa_n^0 + T\kappa_n^c, \quad f_n = f_n^0 + Tf_n^c. 
\end{align}
Here $\Delta^0_{ij} f^{0j}_n = \kappa^0_n f^0_{n,i}$. Using first-order perturbation theory, we have,
\begin{align}
    & \kappa^{c}_n = \int d^4x \sqrt{g^0}\ f^{0i}_n(x)\ \Delta^{(c)}_{ij}\ f^{0j}_n(x)  \\
    & f_n^{ci} (x) = \sum_{m\neq n}\frac{1}{\kappa^0_n - \kappa^0_m}\left(\int d^4x' \sqrt{g^0}\ f^{0j}_m(x')\ \Delta^{(c)}_{jk}\ f^{0k}_n(x')\right)\ f_m^{0i} (x) \label{eigen-corr}
\end{align}
The heat kernel is again defined as \eqref{heat-boson} for bosonic fluctuations and as \eqref{heat-ferm} for fermionic fluctuations using these eigenvalues and eigenfunctions\footnote{Although we will never explicitly require the eigenfunction corrections.}. \\

\noindent Our goal would be to extract the logarithmic corrections in $Q$ and $T$ since they bring different scales to the solution. For this, we again consider the non-zero and zero modes contributions separately. Further, we have two kinds of non-zero modes, namely those that have have non-zero and slightly non-zero eigenvalues. The proper non-zero modes are the ones that are also non-zero on the extremal background. The proper zero modes are the ones that remain zero modes on the near-extremal background. The contributions are schematically divided into three parts,
\begin{align}
    \log Z = \log Z_{\text{nz}} + \log Z_{\text{snz}} + \log Z_{\text{zm}}.
\end{align}
The treatment of the proper non-zero modes will be done in the same way as the non-zero modes of the extremal black hole, since these eigenvalues still scale like $1/Q^2$. As discussed in \cite{Banerjee:2023quv}, these modes give rise to $\log{Q}$ terms only and the temperature dependence in $\log Z$ is polynomially suppressed. The coefficient of logarithmic contribution will also remain the same as the non-zero part of the extremal computation. The same is true for the proper zero modes, which give rise to only $\log Q$ corrections through the path integral measure. Hence, the coefficient of $\log Q$ terms coming from the proper non-zero and proper zero modes will remain equal to their counterparts in the extremal case.\\

\subsubsection*{Bosonic slightly non-zero sector}
\noindent  We will consider the contribution of the slightly non-zero modes separately since these modes will give rise to both $\log Q$ and $\log T$ contributions. Although the proper non-zero and proper zero modes computations are quite similar for bosonic and fermionic fluctuations, the fermionic slightly non-zero modes have important differences from the bosonic ones. Hence, first we will consider the bosonic sector and deal with the fermionic sector later. The slightly non-zero modes were zero modes on the extremal background but got lifted in presence of temperature. Evidently, the eigenvalues of these modes are $\mathcal{O}(T)$ i.e. there is no $\mathcal{O}(1)$ piece. The relevant contribution in the partition function is given as,
\begin{align}
    \log Z_{\text{snz}} = \frac{1}{2}\int_{\epsilon}^{\infty}\frac{ds}{s}\left( \overline{\sum}_ne^{-T\kappa_n^c s} \right). \label{slightly-non-zero}
\end{align}
We introduce the notation $\overline{\sum}_n$ to indicate that we are summing over the zero modes of the extremal solution, which get promoted to slightly non-zero modes. From dimensional analysis, we find that,
\begin{equation}
    \kappa_n^c = \frac{\hat{\kappa}_n^c}{Q},
\end{equation}
where hat denotes a dimensionless object. Similar to the earlier case, we make the transformation $s \to \hat{s} = \frac{Ts}{Q}$ such that,
\begin{align}
    \log Z_{\text{snz}} = \frac{1}{2}\int_{T\epsilon/Q}^{\infty}\frac{d\hat{s}}{\hat{s}}\left(\overline{\sum}_ne^{-\hat{\kappa}_n^c \hat{s}}\right). \label{rescaled-lower-lim}
\end{align}
We can recognize that again the logarithmic corrections arise from the small-$\hat{s}$ integration of the $\hat{s}$-independent part of the term in parentheses. Thus we get,
\begin{equation}
    \log Z_{\text{snz}} = -\frac{1}{2}N_{\text{snz}}\log\left(\frac{T \epsilon}{Q}\right), \quad N_{\text{snz}} \equiv \overline{\sum}_n 1. \label{logT}
\end{equation}
We conclude that the coefficient of this correction is given by the number of zero modes\footnote{We have used the fact that the corrected eigenfunctions are orthonormal at order $T$ and the orthonormality condition is exactly equal to that of the extremal eigenfunctions.} of the extremal solution, which get promoted to non-zero modes when we switch on a small temperature. In terms of the trace of the heat kernel, we find that,
\begin{align}
    N_{\text{snz}} = 8\pi^2 Q^4 (\cosh{\eta_0} - 1)\Bar{K}(0). 
\end{align}
The bar again denotes a sum over the slightly non-zero modes. Since we are considering the quantization problem as finding the spectrum of a modified operator on the extremal background, the boundary cutoff-dependent piece can again be dropped off. Hence, the logarithmic corrections from the cutoff-independent part are given as,
\begin{align}
    \log Z_{\text{snz}} \sim 4\pi^2 Q^4\Bar{K}(0)\log\left(\frac{T \epsilon}{Q}\right).
\end{align}

\subsubsection*{Fermionic slightly non-zero sector}
\noindent Now we consider the fermionic slightly non-zero modes. As mentioned earlier, the treatment of proper non-zero and proper zero modes will still remain the same.  Although these modes have non-zero eigenvalues, we will show that we cannot treat them in a way similar to the bosonic slightly non-zero modes by considering the squared kinetic operator. Instead, we will need to work with the linear derivative kinetic operator itself. To understand the issue better, let us consider the generic form of the eigenvalues of the fermionic kinetic operator, given as\footnote{For slightly non-zero modes, we are denoting the eigenvalues of the kinetic operator with $\kappa$. It is not to be confused with the eigenvalues of the squared operator.} $\kappa_n^f = A + BT$. For slightly non-zero modes, the $\mathcal{O}(1)$ piece is zero i.e. $A = 0$. Thus the eigenvalues of the squared operator have no $\mathcal{O}(T)$ piece since that is proportional to the $\mathcal{O}(1)$ piece itself. Hence the squared kinetic operator has eigenvalues of $\mathcal{O}(T^2)$. Although we could work with these squared eigenvalues and later take their square root, it would be convenient to work with the eigenvalues of the fermionic kinetic operator and not its square while dealing with these slightly non-zero modes. The contribution of these modes to the partition function is given by,
\begin{align}
    \log Z^f_{\text{snz}} = \frac{1}{2}\overline{\sum}_n\log{\kappa}^f_n. 
\end{align}
We again bring in an auxiliary variable to express this like \eqref{schwinger}. However, this variable has different dimensions than that of the variable used for the bosonic case. Thus we have,
\begin{equation}
    \log Z^f_{\text{snz}} = -\frac{1}{2}\int_{\epsilon '}^{\infty}\frac{ds}{s}\left( \overline{\sum}_ne^{-T\kappa_n^c s} \right), 
\end{equation}
where $\epsilon'$ is again a small cutoff which is related to the UV cutoff of the theory. $\kappa_n^c $ now denotes the shifts in the eigenvalues of the fermionic kinetic operator (and not its square, to be emphatic) corresponding to the slightly non-zero modes. These corrections are dimensionless since the eigenvalues of the fermionic kinetic operator scale as
\begin{equation}
    \kappa_n \sim \frac{1}{Q},
\end{equation}
by virtue of it being a one-derivative operator. To extract the logarithmic contributions, we perform the rescaling $s \to \hat{s} = Ts$ so that,
\begin{equation}
   \log Z^f_{\text{snz}} = -\frac{1}{2}\int_{T\epsilon '}^{\infty}\frac{d\hat{s}}{\hat{s}}\left( \overline{\sum}_ne^{-\kappa_n^c \hat{s}} \right) = \frac{1}{2}N^f_{\text{snz}}\log\left(T \epsilon'\right).
\end{equation}
Here, $N^f_{\text{snz}}$ denotes the number of fermionic slightly non-zero modes, defined through \eqref{ferm-no.} and summing over the appropriate modes. It can be easily seen that if we used the eigenvalues of the squared operator, the lower limit of the $s$-integral would be given by $T^2\epsilon$, where $\epsilon$ is the UV cutoff. This is because the eigenvalues of the squared operator\footnote{This should be thought of as a formal square and not a second-order perturbation theory computation.} scale like $1/Q^2$. From this comparison, we find that $\epsilon'\sim\sqrt{\epsilon}$. Thus we do not get any $\log Q$ dependence from this sector.\\

\noindent Now, we will write down the complete logarithmic contributions in charge and temperature. The only distinction between extremal and near-extremal results appears from the slightly non-zero modes sector. On the extremal background, these modes would have contributed like ordinary zero modes where the logarithmic contribution comes from the path integral measure. To make a clean comparison, we add and subtract these contributions of the strictly extremal background. Then the logarithmic contribution in the logarithm of partition function can be written as, 
\begin{align}
    \log Z \sim &\left[c_0 + 8\pi^2Q^4\sum_{r\in \text{snz}}\left(\beta_r - \frac{1}{2}\right)K^r(0)\right]\log Q + 8\pi^2Q^4\sum_{r\in \text{snz}}\frac{1}{2}K^r(0)\log T \nonumber \\
    + & \left[c^f_0 + 8\pi^2Q^4\sum_{f\in \text{snz}}\beta_f\ K^f(0)\right]\log Q + 8\pi^2Q^4\sum_{f\in \text{snz}}K^f(0)\log T. \label{log-NE}
\end{align}
We have explicitly brought out the difference in the coefficients of $\log Q$ terms for the extremal and the near-extremal black holes to draw a proper analogy. Here $r,f$ label different types of bosonic and fermionic slightly non-zero modes respectively. As discussed earlier, $\beta_r$ and $\beta_f$ are the numbers determining the logarithmic contribution for these modes for zero temperature i.e. when these are zero modes. This equation \eqref{log-NE} is one of the main results of this paper and can be applied to compute the near-extremal logarithmic corrections in arbitrary theories of gravity, an analysis which was so far missing in the literature.

\subsection{Regime of validity}
\noindent We analyze the temperature regime where the formula \eqref{log-NE} holds. Since we are considering a near-extremal black hole, we already imposed an upper bound on the temperature as $QT\ll 1$. In this section, we show that there is also a lower bound in temperature till which the final near-extremal partition function is valid.\\

\noindent As discussed earlier, the non-zero and zero mode contributions for a near-extremal black hole can be evaluated similarly to those of an extremal black hole. The difference appears due to the presence of slightly non-zero modes. We want to understand the lowest temperature, till which we can consider these modes to be `slightly' non-zero and below which the currently available methods start treating them as zero modes. Intuitively, this lowest temperature will signify a point where the Gaussian integral flattens enough that it can no longer be treated as a suppressed integral. For a simple Gaussian integral, this regime can be thought of as a limit where the width of the Gaussian becomes much larger than the domain of the integration. To understand this in the heat kernel formalism, let us note that in the identity \eqref{schwinger}, the parameter $A$ should be much larger than the cutoff $\epsilon$. If $A$ becomes comparable with $\epsilon$, we could expand the integrand in a series in $\epsilon$. Under an $\epsilon\to 0$ limit, the integral would be typically divergent depending on the cutoff only, and the logarithmic dependence on the parameter $A$ will be lost. This is equivalent to the flattening of the Gaussian. Similar arguments hold for the partition function computation, where $A$ is replaced by appropriate black hole parameters. This behavior sets the lower bound on temperature through \eqref{rescaled-lower-lim} such that it does not become comparable with the UV cutoff of the theory. Hence, our result\footnote{In such a small scale of temperature, the nonperturbative corrections to the path integral might become important as suggested in \cite{Iliesiu:2022onk}. These effects are not completely understood yet and out of the scope of our current perturbative computations.} \eqref{log-NE} is valid in the following temperature range,
\begin{align}
    y\left(\frac{\epsilon}{Q^2}\right)\ll QT \ll 1. \label{temp-regime}
\end{align}

\noindent Here, $y$ can be a certain power of the dimensionless combination $\frac{\epsilon}{Q^2}$. For different kinds of field fluctuations, the explicit power law dependence might change. Thus the lower limit should be the largest of all these bounds. However, the temperature must be much above this bound. Below this temperature bound, we should start treating these modes as zero modes again in the heat kernel prescription. This indicates that the current machinery is not enough fine-tuned to identify a slightly non-zero mode in this regime. Thus in such a low temperature range, the extremal black hole result \eqref{logQ-coeff-ext} tends to hold.

\subsection{Extremal limit}\label{ext-lim}
\noindent Let us consider a near-extremal black hole in the temperature regime where we can distinguish between zero and slightly non-zero modes. We have shown that a near-extremal and an extremal black hole having the same charges may have different $\log Q$ coefficients, which is a physically consistent observation. Now we will show that a systematic extremal limit of the near-extremal computations is possible and under this limit, we can recover the $\log Q$ coefficient as predicted from the computations on the extremal geometry. \\

\noindent Let us consider the example of a simple Gaussian integral to illustrate the limit,
\begin{align}
    I_{\alpha} = \int_{-\infty}^{\infty} \alpha dx\ \e{-b\alpha^2x^2} = \frac{\sqrt{\pi}}{\sqrt{b}}.
\end{align}
We want to match the values of $I_{\alpha}$ under independent $b\to 0$ limits on both sides of the equation. In particular, we would like to compare the $\alpha$-dependence of the integral\footnote{We have introduced the explicit scaling factors of $\alpha$ to point out that the result is independent of any rescaling of the variable $x$. This feature will be important for slightly non-zero modes.}. Naively it seems that the right hand side is (i.e. after performing the Gaussian integral) independent of $\alpha$. However if we take $b = 0$ on left hand side, the integral seems to pick up a factor of $\alpha$. Thus, there seems to be an apparent discrepancy in the result. Here we show that it is possible to extract this factor of $\alpha$ even after performing the integral. For this, we need to regulate the integral by evaluating it on an interval instead and the divergences will be coming from the length of this interval. We consider the Gaussian error function, defined as,
\begin{align}
    \text{erf}(y) = \frac{2}{\sqrt{\pi}}\int_0^y \text{e}^{-x^2}dx.
\end{align}
Evaluating $I_{\alpha}$ on an interval gives,
    \begin{align}
        I_{\alpha} = \int_{-L}^{L} \alpha dx\ \text{e}^{-b\alpha^2 x^2} = \frac{\sqrt{\pi}}{\sqrt{b}}\ \text{erf}(\sqrt{b}\alpha L). \label{erf}
    \end{align}
If we take a $b \rightarrow 0$ limit on the left side (i.e. in the integrand), we get,
\begin{align}
    I_\alpha = \int \alpha dx = \alpha\cdot 2L.
\end{align}
The $b \to 0$ limit on the right side (i.e. after performing the integral) amounts to studying the behavior of the error function under $b\to 0$ and $L\to\infty$ limits, which gives
\begin{align}
    I_\alpha = \alpha\cdot 2L - \frac{2}{3}L^3\alpha^3 b + \mathcal{O}(b^2).
\end{align}
Hence, in a strict $b \to 0$ limit, both sides agree on the dependence of $\alpha$. A similar analysis of such a limit can be performed for fermionic Gaussian integrals, where the results are not divergent. Instead, the result goes to zero when the integrand is 1, as is typical for Grassmann variables. We will discuss the scaling dependence in the next few lines, skipping an explicit derivation. In presence of a Gaussian suppression, any arbitrary scaling of the integration variable keeps the result unchanged. But if we take the Gaussian exponent to zero, the scaling factor in the measure can no longer be canceled from the Gaussian.\\

\noindent 
These features of Gaussian integrals are crucial in understanding a zero temperature limit of the integral over the slightly non-zero modes. As discussed earlier, the zero modes of an extremal black hole are generated by large gauge transformation parameters that are independent of the black hole scale. To understand the $\log{Q}$ contribution coming from these zero modes, a change of variable is performed from the fields to the gauge parameters \cite{Sen:2012kpz}. The gauge parameters being $Q$-independent, the measure acquires a factor of $Q^{\beta_r}$, coming from the Jacobian of change of variable. The number $\beta_r$ depends on the type of zero mode in consideration. This factor nontrivially contributes since there is no Gaussian suppression for zero modes on the extremal background. However, as discussed above, if there is a Gaussian suppression, the result of the integral is independent of the scale factor. To understand this, we draw an analogy with the integral $I_\alpha$, where $Q^{\beta_r}$ is playing the role of $\alpha$ in our case. Hence, for slightly non-zero modes integral, this normalization does not affect the Gaussian integral result. Since we want to consider an extremal limit, we would first like to make the change of variable to the gauge parameters so that integration limits become $Q$-independent. Then for each slightly non-zero mode integral, we will get a factor of $Q^{\beta_r}$ in the measure and a factor of $Q^{2\beta_r}$ in the Gaussian exponent. This does not change the result of the Gaussian integral when $T\neq 0$. But when we take the $T = 0$ limit, our analysis shows that we get a factor of $Q^{\beta_r}$ coming from each mode, resulting in a coefficient:
    \begin{align}
        Q^{\beta_r(\text{Number of modes})}.
    \end{align}
This precisely agrees with the computations on an extremal black hole, where these modes were treated as zero modes. The argument for fermionic integrals also goes through similarly. \\

\noindent This simple analysis indicates that if we consider the result of Gaussian integral of slightly non-zero modes in the presence of temperature and take a zero temperature limit of the result, it should boil down to the extremal result. The caveat here is that we can only take such a limit insofar as we have not made use of any techniques that necessarily require the temperature of our system to be non-zero. For instance, even when the temperature is non-zero, the contributions coming from an infinite number of slightly non-zero modes typically give divergent factors. To regulate this divergence, a Zeta function regularization was performed in \cite{Banerjee:2023quv}, which finally gave a non-zero $\log T$ contribution. Similar regularization is considered even in the one-loop partition function of the Schwarzian theory \cite{Saad:2019lba}, which in turn is used to draw the conclusions in \cite{Iliesiu:2020qvm, Iliesiu:2022onk}. This regularization explicitly uses that $T \neq 0$, however small it is. Hence, it is not correct to take an extremal limit beyond this regularization. In the heat kernel approach, we cannot take a zero temperature limit after we have extracted the scale dependence of the slightly non-zero eigenvalues in \eqref{rescaled-lower-lim}, which in turn changes the lower limit of the $s$-integral. After this rescaling, it is not possible to take a strict $T = 0$ limit. The limits of applicability of our analysis was discussed in the earlier section.\\

\noindent The main takeaway point of this analysis is that it is possible to recover the extremal $\log Q$ correction from the near-extremal result in any arbitrary theory as long as we do not explicitly assume $T\neq 0$, in performing some kind of regularization. Hence, the near-extremal results are perfectly consistent with the extremal black hole result, which are indeed the complete logarithmic corrections at extremality. This agreement is based on an appropriate extremal limit and after the regularization procedures, it is not possible to smoothly go to extremality. Our work provides a systematic computation of logarithmic corrections to near-extremal entropy, which may be different than the corresponding extremal results.

\section{Pure $\mathcal{N} = 2$ supergravity}\label{N=2}

\noindent In this section, we will compute the logarithmic corrections for a near-extremal black hole in pure $\mathcal{N} = 2$ supergravity theory in four dimensions using the formula \eqref{log-NE}. Our task will be to identify the slightly non-zero modes by analyzing the structure of the near-extremal background, without ever explicitly computing the eigenvalues. The bosonic sector of this theory is the Einstein-Maxwell theory. Hence, the non-rotating black hole solution is described by Reissner-N\"{o}rdstrom geometry. We will first understand the bosonic sector contributions. 

\subsection{Bosonic sector: Einstein-Maxwell theory}\label{EM}
\noindent We need to understand which of the bosonic zero modes of the extremal black hole will get lifted in the presence of temperature. As mentioned earlier, these zero modes are associated with the spontaneous breaking of certain asymptotic symmetries near the AdS$_2$ boundary. To be precise, the $SL(2,R)$ global isometry enhances to the asymptotic symmetries of AdS$_2$, which are large diffeomorphisms forming the Virasoro symmetry algebra. The $U(1)$ and $SO(3)$ symmetries enhance to large gauge transformations near the boundary. These asymptotic symmetries are spontaneously broken to the global part, which is the isometry group of the extremal solution. Hence, we have infinite number of zero modes associated with this breaking. These modes are the tensor modes, $l = 0$ vector modes and $l = 1$ vector modes \cite{Banerjee:2010qc, Banerjee:2011jp, Sen:2012kpz}. Now we will discuss their fate in presence of small temperature correction to the background.

\subsubsection*{Tensor modes on near-extremal background}

\noindent As discussed above, the tensor modes are associated with large diffeomorphisms on AdS$_2$. These fluctuations are asymptotically AdS$_2$, following a particular falloff behavior near the boundary located at large radial distance $\eta = \eta_0$. As a result, the corresponding Ricci scalars are well approximated to the negative constant value of the extremal AdS$_2$ geometry. In fact, deviations from this value are much suppressed in the radial coordiante $\text{e}^{-\eta}$. The large diffeomorphisms preserve these boundary behaviors, so that they are asymptotic symmetries.\\

\noindent The near-extremal background is not asymptotically AdS$_2$, which can be realized from the expression of Ricci scalar,
\begin{align}
    R = -\frac{2}{Q^2} + \frac{24\pi T}{Q}\cosh{\eta} + \mathcal{O}(T)^2.
\end{align}
In fact, near the boundary the linear-order temperature correction grows. Hence, the asymptotic symmetries are lost in presence of temperature. As a result, there will be no zero modes in this sector and we conclude that the tensor modes will become slightly non-zero modes on the near-extremal background. This can be thought of as a consequence of the way the near-extremal solution was constructed by introducing additional mass above the extremal mass.

\subsubsection*{$U(1)$ vector modes on near-extremal background}

\noindent These fluctuations are generated by large gauge transformations, which are symmetries on the extremal background. Since these modes are pure gauge, their corresponding field strengths are zero. Therefore, the flux in presence of these fluctuations is equal to the flux of the extremal background i.e. proportional to the electric charge of the extremal black hole. While constructing the near-extremal solution, we kept the electric charge fixed. From the near-horizon perspective, we find that the flux of the near-extremal solution is still equal to the flux of the extremal solution,
\begin{align}
    F_{U(1)} = \int d\psi d\varphi \sqrt{g} F^{\eta\theta} \propto Q + \mathcal{O}(T^2) 
\end{align}
Note that, if we had considered the near-horizon geometry till deviations of order $T^2$, the flux should be corrected at $\mathcal{O}(T^3)$ since the charge was kept fixed in the full solution.  \\

\noindent Therefore we find that there is an infinite number of field configurations having the same charge as the extremal and near-extremal background. This implies that the large $U(1)$ gauge transformations are still symmetries of the background. Hence, the associated modes are still zero modes.

\subsubsection*{$SO(3)$ vector modes on near-extremal background}

\noindent To understand the fate of the cross-component fluctuations $h_{\mu i}$, we will first put them into the canonical form of Kaluza-Klein ansatz described in \cite{Iliesiu:2020qvm}. Here $\mu$ refers to the indices of AdS$_2$ and $i$ labels the $S^2$ indices. The form of these fluctuations is given by,
\begin{align}
    h_{\mu i} \equiv v_{\mu m}\xi^m_{i} = \partial_{\mu}\Phi\ \xi^m_{i}.
\end{align}
Here $\xi^m_{i}$ are the vectors on S$^2$ that generate $SO(3)$ algebra for $m = 0,\pm 1$. $\Phi$ is the large gauge transformation parameter having a discrete label, that we have suppressed here. On a spherically symmetric background, these fluctuations can be expressed as,
\begin{align}
    ds^2 &= g_{\mu\nu}dx^{\mu}dx^{\nu} + g_{ij}dx^{i}dx^{j} + 2h_{\mu i}dx^{\mu}dx^{j} \\
    &= (g_{\mu\nu} - X(\eta)\partial_{\mu}\Phi \partial_{\nu} \Phi)dx^{\mu}dx^{\nu} + g_{ij}(dx^i + V^m\xi_m^i)(dx^j + V^n\xi_n^j)
\end{align}
$X(\eta) = \sum_{m,p}g_{ij}\xi^i_m \xi^j_p$ can be computed from the knowledge of the background. The reason behind considering this ansatz is that now the fields $V = V^m \xi_m \equiv v^m_{\mu}dx^{\mu}\xi^i_m\partial_i$ can be realized as $SO(3)$-valued gauge fields living on a 2D geometry specified by metric $\Tilde{g}_{\mu\nu} = g_{\mu\nu} - X(\eta)\partial_{\mu}\Phi \partial_{\nu}\Phi$.\\

\noindent The field strength $H = dV - V\wedge V$ corresponding to the $SO(3)$ gauge field is again zero. Hence in presence of these fluctuations, the value of flux should not change from the value of flux of the spherically symmetric background. The question now boils down to whether the modified metric $\Tilde{g}$ can be well approximated to the actual background $g$ from an asymptotic point of view. Analyzing the Ricci scalar, it can be shown that when the background $g$ is extremal, the modified 2D metric is still asymptotically AdS$_2$. Hence in this way, it can be argued that these fluctuations correspond to certain asymptotic symmetries. But when the background is near-extremal, the asymptotic AdS$_2$ structure is lost, as shown earlier. Therefore, this analysis shows that the fluctuations generated by large $SO(3)$ transformations are no longer symmetries in presence of temperature and there are no associated zero modes. The $l = 1$ vector modes are then expected to be slightly non-zero modes on the near-extremal background.\\

\noindent Therefore, we have shown that out of the extremal bosonic zero modes, only the tensor and $l = 1$ vector modes i.e. all the metric zero modes get promoted to slightly non-zero modes on the near-extremal background. The $U(1)$ vector modes are still zero modes of the black hole. This is in agreement with the direct eigenvalue computations of \cite{Banerjee:2023quv}. This result can be traced back to the construction of the near-extremal solution by adding a small mass above extremality while the electric charge was held fixed. Therefore, whenever we keep the charges corresponding to the gauge fields fixed, the corresponding zero modes do not get lifted even in the presence of temperature.

\subsubsection*{Coefficient of logarithmic terms in Einstein-Maxwell sector}
\noindent Since the metric zero modes are lifted, we would like to compute the corresponding trace of the heat kernel which in turn gives the number of such modes. As shown in \cite{Sen:2012kpz}, for these modes $\beta_m = 2$, where we are denoting the zero mode label $r$ by $m$ to indicate metric zero modes. The trace formula gives 
\begin{align}
    K^m(0) = \frac{3}{4\pi^2Q^4}.
\end{align}
Using \eqref{log-NE}, we find the logarithmic corrections coming from the bosonic sector to be,
\begin{align}
    \log Z \sim (c_0 + 9)\log Q + 3\log T. \label{log-N2-bos}
\end{align}
This dependence exactly matches with the explicit eigenvalue computations of \cite{Banerjee:2023quv}.

\subsection{Full logarithmic corrections}
\noindent The extremal black hole solution is half-BPS i.e. it preserves half of the supersymmetry. The preserved supersymmetry gets an infinite extension near the boundary, which in turn results in an infinite number of Goldstone modes. Since the near-extremal black hole has mass above the BPS limit, it does not preserve the supersymmetry that was earlier present in the extremal background. This uplifts the degeneracy of the zero modes. Hence we consider their contribution in the slightly non-zero part. For these modes, $\beta_f = 3$ and the trace of the heat kernel is given as,
\begin{align}
    K^f(0) = -\frac{1}{2\pi^2Q^2}.
\end{align}
The fermionic contribution to the logarithmic corrections is given from \eqref{log-NE},
\begin{align}
    \log Z \sim (c^f_0 - 12)\log Q -4\log T. \label{log-N2-ferm}
\end{align}
Combining the contributions of \eqref{log-N2-bos} and \eqref{log-N2-ferm}, we get the full logarithmic corrections for a near-extremal black hole in pure $\mathcal{N} = 2$ supergravity theory as follows,
 \begin{align}
    \log Z \sim (c_{\text{ext}} - 3)\log Q - \log T. \label{res}
\end{align}
Here $c_{\text{ext}} = c_0 + c^f_0$ is the coefficient of the logarithmic term for the extremal black hole.\\

\noindent Within the regime of validity of the near-extremal partition function, the density of states can be computed from the inverse Laplace transform of the partition function,
\begin{align}
    \rho(E) = \e{S_0}\int d\beta \e{\beta E + \frac{S_1}{\beta}}Z(\beta). 
\end{align}
Here, we have included the saddle point contribution to the partition function, where $S_0 = 16\pi^2Q^2$ denotes the extremal part and $S_1 = 32\pi^3Q^3$ signifies the near-extremal correction. The saddle point contribution can be computed using the standard Gibbons-Hawking-York prescription as described in \cite{Banerjee:2023quv}. $E$ denotes the energy above extremality. $Z(\beta)$ denotes the contribution to the one-loop partition function coming from the logarithmic corrections. Substituting $Z(\beta) = Q^{c_{\text{ext}}-3}\beta$, we get the expression for the density of states as,
\begin{align}
    \rho(E) \sim Q^{c_{\text{ext}}+3} \leftindex_{0}F_1 (3;32\pi^3 Q^3).
\end{align}
The small $E$ expansion of the same goes as, 
\begin{equation}
    \rho(E) \sim \e{S_0}  Q^{c_{\text{ext}}+3}\left[ 1 + \frac{32}{3}(\pi Q )^3 E + \mathcal{O}(E^2 ) \right]. \label{log}
\end{equation}
Hence, the logarithmic correction to entropy will be given by the leading term, whereas the subleading terms will give a polynomial contribution. For the $\mathcal{N} = 2$ theory, the half-BPS extremal solution has a coefficient $c_{\text{ext}} = \frac{23}{12}$ for the logarithmic contribution \cite{Sen:2012kpz}. Thus for the near-extremal solution that we are considering, we have $Z(\beta) \sim Q^{-\frac{13}{12}}\beta$. Corresponding to this, we obtain the density of states as follows.
\begin{align}
    \rho(E) \sim \e{S_0} Q^{\frac{59}{12}}\left[ 1 + \frac{32}{3}(\pi Q )^3 E + \mathcal{O}(E^2 ) \right]. 
\end{align}
This result is valid in the following regime of energy above extremality owing to \eqref{temp-regime},
\begin{align}
    y\left(\frac{\epsilon}{Q^2}\right)\ll QE \ll 1.
\end{align}
Thus the $\log Q$ correction to the entropy of a near-extremal, non-rotating black hole in $\mathcal{N} = 2$ supergravity is given by,
\begin{equation}
    \Delta S_{\log} = \frac{59}{12}\log Q.
\end{equation}
Our result \eqref{res} agrees with the findings of \cite{Iliesiu:2022onk} for a particular saddle point.
\section{$\mathcal{N}=4, 8$ Supergravity}\label{N=4,8}

\noindent In this section, we will discuss the logarithmic corrections to the entropy of non-rotating, near-extremal black holes in $\mathcal{N} = 4, 8$ supergravity theories. We will consider the near-extremal solutions to be small temperature deviations of the $\frac{1}{4}$-BPS and $\frac{1}{8}$-BPS solutions. As discussed earlier, the extremal solutions are parametrized by multiple charges, unlike the $\mathcal{N} = 2$ case. All these charges scale uniformly with the horizon size $a$. Thus we will consider a near-extremal solution having temperature such that $a T\ll 1$. In this temperature regime, the near-horizon geometry can be expressed as in \eqref{gen-NE-NH}.\\

\noindent Hence, the analysis of section \ref{heat-kernel} can be carried out similarly by replacing the charge parameter $Q$ by the extremal horizon size $a$. As discussed, the coefficient of $\log a$ contribution from fully non-zero and fully zero modes should remain the same as the corresponding extremal results presented in \cite{Banerjee:2010qc, Banerjee:2011jp}. In order to find the complete logarithmic contribution, we focus on the slightly non-zero mode sector.\\

\noindent We begin with the zero modes of the extremal solutions in these theories. The $\mathcal{N} = 4$ theory consists of gravity and matter multiplets. The matter multiplet is comprised of $U(1)$ gauge fields, scalars and spin-$1/2$ fields. Thus zero modes only appear from the gauge fields \cite{Banerjee:2010qc}. The gravity multiplet has metric and Rarita-Schwinger fields which coincide with the $\mathcal{N} = 2$ theory field content such that the zero mode structure of this sector is also identical. In addition, there are some $U(1)$ gauge fields that give rise to zero modes \cite{Banerjee:2011jp}. The $\mathcal{N} = 8$ theory contains additional gauge fields and scalars compared to the $\mathcal{N} = 4$ field content \cite{Banerjee:2011jp}. \\ 

\noindent Therefore, the considerations of \cite{Banerjee:2010qc, Banerjee:2011jp} show that the metric and fermionic zero mode spectra of $\mathcal{N} = 4,8$ theories are identical to that of the $\mathcal{N} = 2$ theory \cite{Sen:2012kpz}. The only difference appears due to additional gauge field zero modes. Since we are keeping the charges corresponding to the gauge fields to be fixed to their extremal values, these gauge field zero modes should remain degenerate even in the presence of a non-zero temperature. Thus although the extremal logarithmic corrections are different in these theories, the slightly non-zero mode contribution remains the same. The density of states, and hence the entropy correction again has the form of \eqref{log}, with a coefficient $c_{\text{ext}}+3$ for the logarithmic correction. Here, $c_{\text{ext}}$ is to be substituted by the appropriate value of logarithmic correction coefficient for the corresponding extremal solutions in $\mathcal{N} = 4,8$ theories \cite{Banerjee:2010qc, Banerjee:2011jp}.

\section{Discussions}\label{concl}

\noindent In this paper, we have constructed a methodology to suitably modify the heat kernel prescription to compute the logarithmic corrections for near-extremal black hole entropy. Since the usual methods for (non)-extremal black holes cannot be directly applied in the near-extreme regime, our work provides an important framework to study the logarithmic corrections for these black holes. Our computation is a close cousin to that of an extremal black hole and there are interesting differences with non-extremal black hole computations even though we have turned on a non-zero temperature. For instance, the non-extremal computations are performed with the full geometry, whereas the near-extremal computations are performed with the near-horizon geometry. Our analysis recasts the quantization problem into finding the eigenvalues of a modified operator on the extremal background itself using first-order perturbation theory. The only subtlety is that the throat is no longer effectively infinite but large. Further due to the decoupling feature of the near-horizon and asymptotic regions, we have a clear notion of near-horizon modes and taking their contribution directly gives the entropy of the near-extremal black hole. For non-extremal black holes, such isolation of near-horizon modes is not possible and one has to appropriately subtract thermal gas contributions to obtain the black hole entropy \cite{Sen:2012dw}. \\

\noindent Unlike (non)-extremal black holes, the parameters of a near-extremal black hole do not scale uniformly, which is a unique feature of near-extremal black holes. Using our method, it is possible to suitably separate the logarithmic contributions coming from inverse temperature and the charges. We show that such a separation depends on appropriately extracting the scale dependence of the contribution to the heat kernel coming from slightly non-zero modes. These modes are zero modes of the extremal black hole that become slightly non-degenerate on the near-extremal background. Thus the logarithmic contributions in near-extremal entropy depend on the appropriate identification of extremal zero modes and understanding their fate as we turn on a small temperature. The extremal zero modes are associated with the spontaneous breaking of asymptotic symmetries, localized near the near-horizon AdS$_2$ boundary and the decoupled asymptotic region does not give rise to additional zero modes. The form of the near-extremal perturbations of the extremal geometry governs the uplift or anomalous breaking of these asymptotic symmetries. The near-extremal solution is obtained by introducing a mass/temperature parameter above extremality, keeping the charges fixed. This results in an uplift of the metric zero modes whereas the large gauge symmetries corresponding to the charges should still remain zero modes. An important point is that the logarithmic contributions computed at the level of first-order perturbation theory are robust. From the discussions of section \ref{near-ext-approach} we note that the modes, which are slightly non-zero at first-order in temperature, only give rise to polynomial in temperature corrections at a higher order of perturbation theory. Apparently, it might seem that the modes which remain zero at first-order might get lifted at a higher order and bring in additional logarithmic contributions. However, this is ruled out by the symmetries. Since the charge(s) of the extremal and near-extremal black holes are completely fixed, the corresponding large gauge symmetries always remain preserved. Thus our results correctly capture the logarithmic corrections in the near-extremal regime. \\

\noindent Another important aspect is that our computations are valid in a regime where the temperature parameter not only has an upper cutoff in terms of the charges, but also a lower limit set by the UV cutoff of the theory. The lower cutoff appears to be a limitation of the current techniques. The method can be applied to compute the logarithmic correction to the entropy of a near-extremal solution in any arbitrary theory of gravity.  \\

\noindent We also show that it is possible to correctly obtain the extremal black hole result as a limit of the near-extremal black hole computation. The limit should be taken systematically at the level of the Gaussian integrals of slightly non-zero fluctuations as long as the near-extremal results are not regularized. Such non-zero temperature regularization is typically performed in (super)-Schwarzian theories. In the heat kernel formalism, the lower limit of the Schwinger parameter integration is rescaled by a non-zero temperature and this limit cannot be set strictly to zero. Such regularizations or rescalings explicitly impose a non-zero temperature condition. Thus it is not appropriate to take a zero temperature limit of the final result. This proves that the logarithmic corrections computed from extremal near-horizon geometry are robust results in any arbitrary theory. Thus there is no apparent contradiction between the extremal and near-extremal logarithmic corrections as long as an ill-defined zero temperature limit of the near-extremal result is not taken. \\

\noindent Our current analysis can be naturally extended to higher dimensional spherically symmetric near-extremal black holes as well. The scaling property of the eigenvalues of the kinetic operator remains the same. Thus the generic expression \eqref{log-NE} for the partition function should remain to hold. Depending on the theories in consideration, the coefficient of the logarithmic correction will change due to the count of slightly non-zero modes. Since the extremal black hole solution always has an AdS$_2$ factor, the large diffeomorphism symmetries will always get lifted in the presence of temperature. The AdS$_2$ factor is present even for rotating black holes such that this argument again goes through. These black holes do not have rotational symmetries at extremality, meaning the only slightly non-zero modes coming from the metric will be the large diffeomorphisms of AdS$_2$. Implementing the logic of section \ref{EM}, we can conclude that the logarithm of temperature correction in the logarithm of partition function for 4D Kerr and Kerr-Newman black holes will have a coefficient $\frac{3}{2}$. This is in agreement with the results found in the recent works \cite{Kapec:2023ruw, Rakic:2023vhv}. \\

\noindent Using the modified heat-kernel method we compute the logarithmic corrections to the entropy for near-extremal, non-rotating black hole solutions in 4D Einstein-Maxwell theory and in $\mathcal{N} = 2,4,8$ supergravity theories. We find a perfect agreement of our result with the existing one in the literature for the Einstein-Maxwell theory. For the supergravity theories, the logarithmic contributions agree with that of \cite{Iliesiu:2020qvm, Iliesiu:2022onk} when they consider only a particular black hole saddle. However as per the results of \cite{Iliesiu:2020qvm, Iliesiu:2022onk}, which is derived from an effective theory prescription of these black holes, the partition function of each saddle diverges at $T = 0$ and they propose to regulate the same by taking contributions from an infinite number of saddle points, that modifies the net logarithmic contributions to the entropy. We essentially differ from them in this interpretation. As elaborated in section \ref{ext-lim}, a zero temperature limit cannot be taken from the regularized near-extremal result. Thus there is no apparent divergence in the result and a regularization is not required. However, it would be interesting to compare the computations of \cite{Iliesiu:2020qvm, Iliesiu:2022onk} with a direct 4D Euclidean action computations vide \cite{Banerjee:2023quv} for the same systems. One may hope to try to understand the implications, if any, of the sum over saddle points of the effective theory in the 4D picture in the context of the black hole solutions under consideration. This could help us resolve the apparent disagreement over interpretation that currently seems to be there. We leave this for our future study.

\acknowledgments

\noindent We are extremely grateful to Ashoke Sen for discussions and his comments on the work. MS would like to thank ICTS, Bengaluru for their hospitality during an important stage of the work. MS would also like to thank Arindam Bhattacharjee and Rajesh Kumar Gupta for useful discussions. The work of NB is partially supported by SERB POWER grant. Finally, we are thankful to the people of India for their generous support towards fundamental research.

\bibliography{NElog.bib}

\providecommand{\href}[2]{#2}\begingroup\raggedright\begin{thebibliography}{10}

\bibitem{Bekenstein:1973ur}
J.~D. Bekenstein, \emph{{Black holes and entropy}},
  \href{https://doi.org/10.1103/PhysRevD.7.2333}{\emph{Phys. Rev. D} {\bfseries
  7} (1973) 2333}.

\bibitem{Hawking:1976de}
S.~W. Hawking, \emph{{Black Holes and Thermodynamics}},
  \href{https://doi.org/10.1103/PhysRevD.13.191}{\emph{Phys. Rev. D} {\bfseries
  13} (1976) 191}.

\bibitem{Solodukhin:1994yz}
S.~N. Solodukhin, \emph{{The Conical singularity and quantum corrections to
  entropy of black hole}},
  \href{https://doi.org/10.1103/PhysRevD.51.609}{\emph{Phys. Rev. D} {\bfseries
  51} (1995) 609} [\href{https://arxiv.org/abs/hep-th/9407001}{{\ttfamily
  hep-th/9407001}}].

\bibitem{Mann:1997hm}
R.~B. Mann and S.~N. Solodukhin, \emph{{Universality of quantum entropy for
  extreme black holes}},
  \href{https://doi.org/10.1016/S0550-3213(98)00094-7}{\emph{Nucl. Phys. B}
  {\bfseries 523} (1998) 293}
  [\href{https://arxiv.org/abs/hep-th/9709064}{{\ttfamily hep-th/9709064}}].

\bibitem{Medved:2004eh}
A.~J.~M. Medved, \emph{{A Comment on black hole entropy or does nature abhor a
  logarithm?}}, \href{https://doi.org/10.1088/0264-9381/22/1/009}{\emph{Class.
  Quant. Grav.} {\bfseries 22} (2005) 133}
  [\href{https://arxiv.org/abs/gr-qc/0406044}{{\ttfamily gr-qc/0406044}}].

\bibitem{Cai:2009ua}
R.-G. Cai, L.-M. Cao and N.~Ohta, \emph{{Black Holes in Gravity with Conformal
  Anomaly and Logarithmic Term in Black Hole Entropy}},
  \href{https://doi.org/10.1007/JHEP04(2010)082}{\emph{JHEP} {\bfseries 04}
  (2010) 082} [\href{https://arxiv.org/abs/0911.4379}{{\ttfamily 0911.4379}}].

\bibitem{Aros:2010jb}
R.~Aros, D.~E. Diaz and A.~Montecinos, \emph{{Logarithmic correction to BH
  entropy as Noether charge}},
  \href{https://doi.org/10.1007/JHEP07(2010)012}{\emph{JHEP} {\bfseries 07}
  (2010) 012} [\href{https://arxiv.org/abs/1003.1083}{{\ttfamily 1003.1083}}].

\bibitem{Gibbons:1976ue}
G.~W. Gibbons and S.~W. Hawking, \emph{{Action Integrals and Partition
  Functions in Quantum Gravity}},
  \href{https://doi.org/10.1103/PhysRevD.15.2752}{\emph{Phys. Rev. D}
  {\bfseries 15} (1977) 2752}.

\bibitem{York:1986it}
J.~W. York, Jr., \emph{{Black hole thermodynamics and the Euclidean Einstein
  action}}, \href{https://doi.org/10.1103/PhysRevD.33.2092}{\emph{Phys. Rev. D}
  {\bfseries 33} (1986) 2092}.

\bibitem{Sen:2005wa}
A.~Sen, \emph{{Black hole entropy function and the attractor mechanism in
  higher derivative gravity}},
  \href{https://doi.org/10.1088/1126-6708/2005/09/038}{\emph{JHEP} {\bfseries
  09} (2005) 038} [\href{https://arxiv.org/abs/hep-th/0506177}{{\ttfamily
  hep-th/0506177}}].

\bibitem{Sen:2008yk}
A.~Sen, \emph{{Entropy Function and AdS(2) / CFT(1) Correspondence}},
  \href{https://doi.org/10.1088/1126-6708/2008/11/075}{\emph{JHEP} {\bfseries
  11} (2008) 075} [\href{https://arxiv.org/abs/0805.0095}{{\ttfamily
  0805.0095}}].

\bibitem{Sen:2007qy}
A.~Sen, \emph{{Black Hole Entropy Function, Attractors and Precision Counting
  of Microstates}}, \href{https://doi.org/10.1007/s10714-008-0626-4}{\emph{Gen.
  Rel. Grav.} {\bfseries 40} (2008) 2249}
  [\href{https://arxiv.org/abs/0708.1270}{{\ttfamily 0708.1270}}].

\bibitem{Sen:2008vm}
A.~Sen, \emph{{Quantum Entropy Function from AdS(2)/CFT(1) Correspondence}},
  \href{https://doi.org/10.1142/S0217751X09045893}{\emph{Int. J. Mod. Phys. A}
  {\bfseries 24} (2009) 4225}
  [\href{https://arxiv.org/abs/0809.3304}{{\ttfamily 0809.3304}}].

\bibitem{Banerjee:2009af}
N.~Banerjee, S.~Banerjee, R.~K. Gupta, I.~Mandal and A.~Sen,
  \emph{{Supersymmetry, Localization and Quantum Entropy Function}},
  \href{https://doi.org/10.1007/JHEP02(2010)091}{\emph{JHEP} {\bfseries 02}
  (2010) 091} [\href{https://arxiv.org/abs/0905.2686}{{\ttfamily 0905.2686}}].

\bibitem{Banerjee:2010qc}
S.~Banerjee, R.~K. Gupta and A.~Sen, \emph{{Logarithmic Corrections to Extremal
  Black Hole Entropy from Quantum Entropy Function}},
  \href{https://doi.org/10.1007/JHEP03(2011)147}{\emph{JHEP} {\bfseries 03}
  (2011) 147} [\href{https://arxiv.org/abs/1005.3044}{{\ttfamily 1005.3044}}].

\bibitem{Karan:2019gyn}
S.~Karan, G.~Banerjee and B.~Panda, \emph{{Seeley-DeWitt Coefficients in
  $\mathcal{N}=2$ Einstein-Maxwell Supergravity Theory and Logarithmic
  Corrections to $\mathcal{N}=2$ Extremal Black Hole Entropy}},
  \href{https://doi.org/10.1007/JHEP08(2019)056}{\emph{JHEP} {\bfseries 08}
  (2019) 056} [\href{https://arxiv.org/abs/1905.13058}{{\ttfamily
  1905.13058}}].

\bibitem{Banerjee:2020wbr}
G.~Banerjee, S.~Karan and B.~Panda, \emph{{Logarithmic correction to the
  entropy of extremal black holes in $ \mathcal{N} $ = 1 Einstein-Maxwell
  supergravity}}, \href{https://doi.org/10.1007/JHEP01(2021)090}{\emph{JHEP}
  {\bfseries 01} (2021) 090}
  [\href{https://arxiv.org/abs/2007.11497}{{\ttfamily 2007.11497}}].

\bibitem{Banerjee:2011jp}
S.~Banerjee, R.~K. Gupta, I.~Mandal and A.~Sen, \emph{{Logarithmic Corrections
  to N=4 and N=8 Black Hole Entropy: A One Loop Test of Quantum Gravity}},
  \href{https://doi.org/10.1007/JHEP11(2011)143}{\emph{JHEP} {\bfseries 11}
  (2011) 143} [\href{https://arxiv.org/abs/1106.0080}{{\ttfamily 1106.0080}}].

\bibitem{Sen:2012cj}
A.~Sen, \emph{{Logarithmic Corrections to Rotating Extremal Black Hole Entropy
  in Four and Five Dimensions}},
  \href{https://doi.org/10.1007/s10714-012-1373-0}{\emph{Gen. Rel. Grav.}
  {\bfseries 44} (2012) 1947}
  [\href{https://arxiv.org/abs/1109.3706}{{\ttfamily 1109.3706}}].

\bibitem{Sen:2012dw}
A.~Sen, \emph{{Logarithmic Corrections to Schwarzschild and Other Non-extremal
  Black Hole Entropy in Different Dimensions}},
  \href{https://doi.org/10.1007/JHEP04(2013)156}{\emph{JHEP} {\bfseries 04}
  (2013) 156} [\href{https://arxiv.org/abs/1205.0971}{{\ttfamily 1205.0971}}].

\bibitem{Bhattacharyya:2012wz}
S.~Bhattacharyya, B.~Panda and A.~Sen, \emph{{Heat Kernel Expansion and
  Extremal Kerr-Newmann Black Hole Entropy in Einstein-Maxwell Theory}},
  \href{https://doi.org/10.1007/JHEP08(2012)084}{\emph{JHEP} {\bfseries 08}
  (2012) 084} [\href{https://arxiv.org/abs/1204.4061}{{\ttfamily 1204.4061}}].

\bibitem{Sen:2012kpz}
A.~Sen, \emph{{Logarithmic Corrections to N=2 Black Hole Entropy: An Infrared
  Window into the Microstates}},
  \href{https://doi.org/10.1007/s10714-012-1336-5}{\emph{Gen. Rel. Grav.}
  {\bfseries 44} (2012) 1207}
  [\href{https://arxiv.org/abs/1108.3842}{{\ttfamily 1108.3842}}].

\bibitem{Karan:2020njm}
S.~Karan and B.~Panda, \emph{{Logarithmic corrections to black hole entropy in
  matter coupled $\mathcal{N} \geq 1$ Einstein-Maxwell supergravity}},
  \href{https://doi.org/10.1007/JHEP05(2021)104}{\emph{JHEP} {\bfseries 05}
  (2021) 104} [\href{https://arxiv.org/abs/2012.12227}{{\ttfamily
  2012.12227}}].

\bibitem{Karan:2021teq}
S.~Karan and B.~Panda, \emph{{Generalized Einstein-Maxwell theory:
  Seeley-DeWitt coefficients and logarithmic corrections to the entropy of
  extremal and nonextremal black holes}},
  \href{https://doi.org/10.1103/PhysRevD.104.046010}{\emph{Phys. Rev. D}
  {\bfseries 104} (2021) 046010}
  [\href{https://arxiv.org/abs/2104.06381}{{\ttfamily 2104.06381}}].

\bibitem{Banerjee:2021pdy}
G.~Banerjee and B.~Panda, \emph{{Logarithmic corrections to the entropy of
  non-extremal black holes in $ \mathcal{N} $ = 1 Einstein-Maxwell
  supergravity}}, \href{https://doi.org/10.1007/JHEP11(2021)214}{\emph{JHEP}
  {\bfseries 11} (2021) 214}
  [\href{https://arxiv.org/abs/2109.04407}{{\ttfamily 2109.04407}}].

\bibitem{Karan:2022dfy}
S.~Karan and G.~S. Punia, \emph{{Logarithmic correction to black hole entropy
  in universal low-energy string theory models}},
  \href{https://doi.org/10.1007/JHEP03(2023)028}{\emph{JHEP} {\bfseries 03}
  (2023) 028} [\href{https://arxiv.org/abs/2210.16230}{{\ttfamily
  2210.16230}}].

\bibitem{Strominger:1996sh}
A.~Strominger and C.~Vafa, \emph{{Microscopic origin of the Bekenstein-Hawking
  entropy}}, \href{https://doi.org/10.1016/0370-2693(96)00345-0}{\emph{Phys.
  Lett. B} {\bfseries 379} (1996) 99}
  [\href{https://arxiv.org/abs/hep-th/9601029}{{\ttfamily hep-th/9601029}}].

\bibitem{Breckenridge:1996is}
J.~C. Breckenridge, R.~C. Myers, A.~W. Peet and C.~Vafa, \emph{{D-branes and
  spinning black holes}},
  \href{https://doi.org/10.1016/S0370-2693(96)01460-8}{\emph{Phys. Lett. B}
  {\bfseries 391} (1997) 93}
  [\href{https://arxiv.org/abs/hep-th/9602065}{{\ttfamily hep-th/9602065}}].

\bibitem{David:2006ru}
J.~R. David, D.~P. Jatkar and A.~Sen, \emph{{Dyon Spectrum in N=4
  Supersymmetric Type II String Theories}},
  \href{https://doi.org/10.1088/1126-6708/2006/11/073}{\emph{JHEP} {\bfseries
  11} (2006) 073} [\href{https://arxiv.org/abs/hep-th/0607155}{{\ttfamily
  hep-th/0607155}}].

\bibitem{David:2006ud}
J.~R. David, D.~P. Jatkar and A.~Sen, \emph{{Dyon spectrum in generic N=4
  supersymmetric Z(N) orbifolds}},
  \href{https://doi.org/10.1088/1126-6708/2007/01/016}{\emph{JHEP} {\bfseries
  01} (2007) 016} [\href{https://arxiv.org/abs/hep-th/0609109}{{\ttfamily
  hep-th/0609109}}].

\bibitem{Gupta:2008ki}
R.~K. Gupta and A.~Sen, \emph{{Ads(3)/CFT(2) to Ads(2)/CFT(1)}},
  \href{https://doi.org/10.1088/1126-6708/2009/04/034}{\emph{JHEP} {\bfseries
  04} (2009) 034} [\href{https://arxiv.org/abs/0806.0053}{{\ttfamily
  0806.0053}}].

\bibitem{Banerjee:2007ub}
N.~Banerjee, D.~P. Jatkar and A.~Sen, \emph{{Adding Charges to N=4 Dyons}},
  \href{https://doi.org/10.1088/1126-6708/2007/07/024}{\emph{JHEP} {\bfseries
  07} (2007) 024} [\href{https://arxiv.org/abs/0705.1433}{{\ttfamily
  0705.1433}}].

\bibitem{Banerjee:2008ky}
N.~Banerjee, D.~P. Jatkar and A.~Sen, \emph{{Asymptotic Expansion of the N=4
  Dyon Degeneracy}},
  \href{https://doi.org/10.1088/1126-6708/2009/05/121}{\emph{JHEP} {\bfseries
  05} (2009) 121} [\href{https://arxiv.org/abs/0810.3472}{{\ttfamily
  0810.3472}}].

\bibitem{Banerjee:2009uk}
N.~Banerjee, I.~Mandal and A.~Sen, \emph{{Black Hole Hair Removal}},
  \href{https://doi.org/10.1088/1126-6708/2009/07/091}{\emph{JHEP} {\bfseries
  07} (2009) 091} [\href{https://arxiv.org/abs/0901.0359}{{\ttfamily
  0901.0359}}].

\bibitem{H:2023qko}
A.~A. H., P.~V. Athira, C.~Chowdhury and A.~Sen, \emph{{Logarithmic Correction
  to BPS Black Hole Entropy from Supersymmetric Index at Finite Temperature}},
  \href{https://arxiv.org/abs/2306.07322}{{\ttfamily 2306.07322}}.

\bibitem{Anupam:2023yns}
A.~H. Anupam, C.~Chowdhury and A.~Sen, \emph{{Revisiting Logarithmic Correction
  to Five Dimensional BPS Black Hole Entropy}},
  \href{https://arxiv.org/abs/2308.00038}{{\ttfamily 2308.00038}}.

\bibitem{Nayak:2018qej}
P.~Nayak, A.~Shukla, R.~M. Soni, S.~P. Trivedi and V.~Vishal, \emph{{On the
  Dynamics of Near-Extremal Black Holes}},
  \href{https://doi.org/10.1007/JHEP09(2018)048}{\emph{JHEP} {\bfseries 09}
  (2018) 048} [\href{https://arxiv.org/abs/1802.09547}{{\ttfamily
  1802.09547}}].

\bibitem{Moitra:2019bub}
U.~Moitra, S.~K. Sake, S.~P. Trivedi and V.~Vishal, \emph{{Jackiw-Teitelboim
  Gravity and Rotating Black Holes}},
  \href{https://doi.org/10.1007/JHEP11(2019)047}{\emph{JHEP} {\bfseries 11}
  (2019) 047} [\href{https://arxiv.org/abs/1905.10378}{{\ttfamily
  1905.10378}}].

\bibitem{Iliesiu:2020qvm}
L.~V. Iliesiu and G.~J. Turiaci, \emph{{The statistical mechanics of
  near-extremal black holes}},
  \href{https://doi.org/10.1007/JHEP05(2021)145}{\emph{JHEP} {\bfseries 05}
  (2021) 145} [\href{https://arxiv.org/abs/2003.02860}{{\ttfamily
  2003.02860}}].

\bibitem{Heydeman:2020hhw}
M.~Heydeman, L.~V. Iliesiu, G.~J. Turiaci and W.~Zhao, \emph{{The statistical
  mechanics of near-BPS black holes}},
  \href{https://doi.org/10.1088/1751-8121/ac3be9}{\emph{J. Phys. A} {\bfseries
  55} (2022) 014004} [\href{https://arxiv.org/abs/2011.01953}{{\ttfamily
  2011.01953}}].

\bibitem{Kolekar:2018sba}
K.~S. Kolekar and K.~Narayan, \emph{{AdS$_2$ dilaton gravity from reductions of
  some nonrelativistic theories}},
  \href{https://doi.org/10.1103/PhysRevD.98.046012}{\emph{Phys. Rev. D}
  {\bfseries 98} (2018) 046012}
  [\href{https://arxiv.org/abs/1803.06827}{{\ttfamily 1803.06827}}].

\bibitem{Banerjee:2021vjy}
N.~Banerjee, T.~Mandal, A.~Rudra and M.~Saha, \emph{{Equivalence of JT gravity
  and near-extremal black hole dynamics in higher derivative theory}},
  \href{https://doi.org/10.1007/JHEP01(2022)124}{\emph{JHEP} {\bfseries 01}
  (2022) 124} [\href{https://arxiv.org/abs/2110.04272}{{\ttfamily
  2110.04272}}].

\bibitem{Bhattacharyya:2023gvg}
A.~Bhattacharyya, S.~Ghosh and S.~Pal, \emph{{Aspects of $T\bar{T}+J\bar{T }$
  deformed 2D topological gravity : from partition function to late-time SFF}},
   \href{https://arxiv.org/abs/2309.16658}{{\ttfamily 2309.16658}}.

\bibitem{Iliesiu:2022onk}
L.~V. Iliesiu, S.~Murthy and G.~J. Turiaci, \emph{{Revisiting the Logarithmic
  Corrections to the Black Hole Entropy}},
  \href{https://arxiv.org/abs/2209.13608}{{\ttfamily 2209.13608}}.

\bibitem{Banerjee:2023quv}
N.~Banerjee and M.~Saha, \emph{{Revisiting leading quantum corrections to near
  extremal black hole thermodynamics}},
  \href{https://doi.org/10.1007/JHEP07(2023)010}{\emph{JHEP} {\bfseries 07}
  (2023) 010} [\href{https://arxiv.org/abs/2303.12415}{{\ttfamily
  2303.12415}}].

\bibitem{Fotopoulos:2019vac}
A.~Fotopoulos, S.~Stieberger, T.~R. Taylor and B.~Zhu, \emph{{Extended BMS
  Algebra of Celestial CFT}},
  \href{https://doi.org/10.1007/JHEP03(2020)130}{\emph{JHEP} {\bfseries 03}
  (2020) 130} [\href{https://arxiv.org/abs/1912.10973}{{\ttfamily
  1912.10973}}].

\bibitem{Fan:2019emx}
W.~Fan, A.~Fotopoulos and T.~R. Taylor, \emph{{Soft Limits of Yang-Mills
  Amplitudes and Conformal Correlators}},
  \href{https://doi.org/10.1007/JHEP05(2019)121}{\emph{JHEP} {\bfseries 05}
  (2019) 121} [\href{https://arxiv.org/abs/1903.01676}{{\ttfamily
  1903.01676}}].

\bibitem{Fotopoulos:2020bqj}
A.~Fotopoulos, S.~Stieberger, T.~R. Taylor and B.~Zhu, \emph{{Extended Super
  BMS Algebra of Celestial CFT}},
  \href{https://doi.org/10.1007/JHEP09(2020)198}{\emph{JHEP} {\bfseries 09}
  (2020) 198} [\href{https://arxiv.org/abs/2007.03785}{{\ttfamily
  2007.03785}}].

\bibitem{Banerjee:2021uxe}
N.~Banerjee, T.~Rahnuma and R.~K. Singh, \emph{{Asymptotic symmetry of four
  dimensional Einstein-Yang-Mills and Einstein-Maxwell theory}},
  \href{https://doi.org/10.1007/JHEP01(2022)033}{\emph{JHEP} {\bfseries 01}
  (2022) 033} [\href{https://arxiv.org/abs/2110.15657}{{\ttfamily
  2110.15657}}].

\bibitem{Banerjee:2022hpo}
N.~Banerjee, T.~Rahnuma and R.~K. Singh, \emph{{Soft and collinear limits in $
  \mathcal{N} $ = 8 supergravity using double copy formalism}},
  \href{https://doi.org/10.1007/JHEP04(2023)126}{\emph{JHEP} {\bfseries 04}
  (2023) 126} [\href{https://arxiv.org/abs/2212.11480}{{\ttfamily
  2212.11480}}].

\bibitem{Banerjee:2022lnz}
N.~Banerjee, T.~Rahnuma and R.~K. Singh, \emph{{Asymptotic Symmetry algebra of
  $\mathcal{N}=8$ Supergravity}},
  \href{https://arxiv.org/abs/2212.12133}{{\ttfamily 2212.12133}}.

\bibitem{Saad:2019lba}
P.~Saad, S.~H. Shenker and D.~Stanford, \emph{{JT gravity as a matrix
  integral}},  \href{https://arxiv.org/abs/1903.11115}{{\ttfamily 1903.11115}}.

\bibitem{Kapec:2023ruw}
D.~Kapec, A.~Sheta, A.~Strominger and C.~Toldo, \emph{{Logarithmic Corrections
  to Kerr Thermodynamics}},  \href{https://arxiv.org/abs/2310.00848}{{\ttfamily
  2310.00848}}.

\bibitem{Rakic:2023vhv}
I.~Rakic, M.~Rangamani and G.~J. Turiaci, \emph{{Thermodynamics of the
  near-extremal Kerr spacetime}},
  \href{https://arxiv.org/abs/2310.04532}{{\ttfamily 2310.04532}}.

\end{thebibliography}\endgroup

\end{document}